\documentclass[twocolumn]{aastex701}
\usepackage{graphicx}
\usepackage{mathcomp}
\graphicspath{{./}{figures/}}

\begin{document}

\title{First Associated Neutrino Search for a Failed Supernova Candidate with Super-Kamiokande}

\newcommand{\AFFicrr}{\affiliation{Kamioka Observatory, Institute for Cosmic Ray Research, University of Tokyo, Kamioka, Gifu 506-1205, Japan}}
\newcommand{\AFFkashiwa}{\affiliation{Research Center for Cosmic Neutrinos, Institute for Cosmic Ray Research, University of Tokyo, Kashiwa, Chiba 277-8582, Japan}}
\newcommand{\AFFipmu}{\affiliation{Kavli Institute for the Physics and
Mathematics of the Universe (WPI), The University of Tokyo Institutes for Advanced Study,
University of Tokyo, Kashiwa, Chiba 277-8583, Japan }}
\newcommand{\AFFmad}{\affiliation{Department of Theoretical Physics, University Autonoma Madrid, 28049 Madrid, Spain}}
\newcommand{\AFFubc}{\affiliation{Department of Physics and Astronomy, University of British Columbia, Vancouver, BC, V6T1Z4, Canada}}
\newcommand{\AFFbu}{\affiliation{Department of Physics, Boston University, Boston, MA 02215, USA}}
\newcommand{\AFFuci}{\affiliation{Department of Physics and Astronomy, University of California, Irvine, Irvine, CA 92697-4575, USA }}
\newcommand{\AFFcsu}{\affiliation{Department of Physics, California State University, Dominguez Hills, Carson, CA 90747, USA}}
\newcommand{\AFFcnm}{\affiliation{Institute for Universe and Elementary Particles, Chonnam National University, Gwangju 61186, Korea}}
\newcommand{\AFFduke}{\affiliation{Department of Physics, Duke University, Durham NC 27708, USA}}
\newcommand{\AFFgifu}{\affiliation{Department of Physics, Gifu University, Gifu, Gifu 501-1193, Japan}}
\newcommand{\AFFgist}{\affiliation{GIST College, Gwangju Institute of Science and Technology, Gwangju 500-712, Korea}}
\newcommand{\AFFuh}{\affiliation{Department of Physics and Astronomy, University of Hawaii, Honolulu, HI 96822, USA}}
\newcommand{\AFFicl}{\affiliation{Department of Physics, Imperial College London , London, SW7 2AZ, United Kingdom }}
\newcommand{\AFFkek}{\affiliation{High Energy Accelerator Research Organization (KEK), Tsukuba, Ibaraki 305-0801, Japan }}
\newcommand{\AFFkobe}{\affiliation{Department of Physics, Kobe University, Kobe, Hyogo 657-8501, Japan}}
\newcommand{\AFFkyoto}{\affiliation{Department of Physics, Kyoto University, Kyoto, Kyoto 606-8502, Japan}}
\newcommand{\AFFliv}{\affiliation{Department of Physics, University of Liverpool, Liverpool, L69 7ZE, United Kingdom}}
\newcommand{\AFFmiyagi}{\affiliation{Department of Physics, Miyagi University of Education, Sendai, Miyagi 980-0845, Japan}}
\newcommand{\AFFnagoya}{\affiliation{Institute for Space-Earth Environmental Research, Nagoya University, Nagoya, Aichi 464-8602, Japan}}
\newcommand{\AFFkmi}{\affiliation{Kobayashi-Maskawa Institute for the Origin of Particles and the Universe, Nagoya University, Nagoya, Aichi 464-8602, Japan}}
\newcommand{\AFFpol}{\affiliation{National Centre For Nuclear Research, 02-093 Warsaw, Poland}}
\newcommand{\AFFsuny}{\affiliation{Department of Physics and Astronomy, State University of New York at Stony Brook, NY 11794-3800, USA}}
\newcommand{\AFFokayama}{\affiliation{Department of Physics, Okayama University, Okayama, Okayama 700-8530, Japan }}
\newcommand{\AFFosaka}{\affiliation{Department of Physics, Osaka University, Toyonaka, Osaka 560-0043, Japan}}
\newcommand{\AFFox}{\affiliation{Department of Physics, Oxford University, Oxford, OX1 3PU, United Kingdom}}
\newcommand{\AFFqmul}{\affiliation{School of Physics and Astronomy, Queen Mary University of London, London, E1 4NS, United Kingdom}}
\newcommand{\AFFregina}{\affiliation{Department of Physics, University of Regina, 3737 Wascana Parkway, Regina, SK, S4SOA2, Canada}}
\newcommand{\AFFseoul}{\affiliation{Department of Physics and Astronomy, Seoul National University, Seoul 151-742, Korea}}
\newcommand{\AFFsheff}{\affiliation{School of Mathematical and Physical Sciences, University of Sheffield, S3 7RH, Sheffield, United Kingdom}}
\newcommand{\AFFshizuokasc}{\affiliation{Department of Informatics in
Social Welfare, Shizuoka University of Welfare, Yaizu, Shizuoka, 425-8611, Japan}}
\newcommand{\AFFstfc}{\affiliation{STFC, Rutherford Appleton Laboratory, Harwell Oxford, and Daresbury Laboratory, Warrington, OX11 0QX, United Kingdom}}
\newcommand{\AFFskk}{\affiliation{Department of Physics, Sungkyunkwan University, Suwon 440-746, Korea}}
\newcommand{\AFFtodai}{\affiliation{Department of Physics, University of Tokyo, Bunkyo, Tokyo 113-0033, Japan }}
\newcommand{\AFFtit}{\affiliation{Department of Physics, Institute of Science Tokyo, Meguro, Tokyo 152-8551, Japan }}
\newcommand{\AFFtus}{\affiliation{Department of Physics and Astronomy, Faculty of Science and Technology, Tokyo University of Science, Noda, Chiba 278-8510, Japan }}
\newcommand{\AFFtriumf}{\affiliation{TRIUMF, 4004 Wesbrook Mall, Vancouver, BC, V6T2A3, Canada }}
\newcommand{\AFFtokai}{\affiliation{Department of Physics, Tokai University, Hiratsuka, Kanagawa 259-1292, Japan}}
\newcommand{\AFFtsinghua}{\affiliation{Department of Engineering Physics, Tsinghua University, Beijing, 100084, China}}
\newcommand{\AFFynu}{\affiliation{Department of Physics, Yokohama National University, Yokohama, Kanagawa, 240-8501, Japan}}
\newcommand{\AFFllr}{\affiliation{Ecole Polytechnique, IN2P3-CNRS, Laboratoire Leprince-Ringuet, F-91120 Palaiseau, France }}
\newcommand{\AFFbari}{\affiliation{ Dipartimento Interuniversitario di Fisica, INFN Sezione di Bari and Universit\`a e Politecnico di Bari, I-70125, Bari, Italy}}
\newcommand{\AFFnapoli}{\affiliation{Dipartimento di Fisica, INFN Sezione di Napoli and Universit\`a di Napoli, I-80126, Napoli, Italy}}
\newcommand{\AFFroma}{\affiliation{INFN Sezione di Roma and Universit\`a di Roma ``La Sapienza'', I-00185, Roma, Italy}}
\newcommand{\AFFpadova}{\affiliation{Dipartimento di Fisica, INFN Sezione di Padova and Universit\`a di Padova, I-35131, Padova, Italy}}
\newcommand{\AFFkeio}{\affiliation{Department of Physics, Keio University, Yokohama, Kanagawa, 223-8522, Japan}}
\newcommand{\AFFwinnipeg}{\affiliation{Department of Physics, University of Winnipeg, MB R3J 3L8, Canada }}
\newcommand{\AFFkcl}{\affiliation{Department of Physics, King's College London, London, WC2R 2LS, UK }}
\newcommand{\AFFwarwick}{\affiliation{Department of Physics, University of Warwick, Coventry, CV4 7AL, UK }}
\newcommand{\AFFral}{\affiliation{Rutherford Appleton Laboratory, Harwell, Oxford, OX11 0QX, UK }}
\newcommand{\AFFwu}{\affiliation{Faculty of Physics, University of Warsaw, Warsaw, 02-093, Poland }}
\newcommand{\AFFbcit}{\affiliation{Department of Physics, British Columbia Institute of Technology, Burnaby, BC, V5G 3H2, Canada }}
\newcommand{\AFFtohoku}{\affiliation{Department of Physics, Faculty of Science, Tohoku University, Sendai, Miyagi, 980-8578, Japan }}
\newcommand{\AFFicise}{\affiliation{Institute For Interdisciplinary Research in Science and Education, ICISE, Quy Nhon, 55121, Vietnam }}
\newcommand{\AFFilance}{\affiliation{ILANCE, CNRS - University of Tokyo International Research Laboratory, Kashiwa, Chiba 277-8582, Japan}}
\newcommand{\AFFibs}{\affiliation{Center for Underground Physics, Institute for Basic Science (IBS), Daejeon, 34126, Korea}}
\newcommand{\AFFglasgow}{\affiliation{School of Physics and Astronomy, University of Glasgow, Glasgow, Scotland, G12 8QQ, United Kingdom}}
\newcommand{\AFFoecu}{\affiliation{Media Communication Center, Osaka Electro-Communication University, Neyagawa, Osaka, 572-8530, Japan}}
\newcommand{\AFFminn}{\affiliation{School of Physics and Astronomy, University of Minnesota, Minneapolis, MN  55455, USA}}
\newcommand{\AFFsilesia}{\affiliation{August Che\l{}kowski Institute of Physics, University of Silesia in Katowice, 75 Pu\l{}ku Piechoty 1, 41-500 Chorz\'{o}w, Poland}}
\newcommand{\AFFtoyama}{\affiliation{Faculty of Science, University of Toyama, Toyama City, Toyama 930-8555, Japan}}
\newcommand{\AFFbmcc}{\affiliation{Science Department, Borough of Manhattan Community College / City University of New York, New York, New York, 1007, USA.}}
\newcommand{\AFFnumazu}{\affiliation{National Institute of Technology, Numazu College, Numazu 410-8501, Japan}}

\AFFicrr
\AFFkashiwa
\AFFmad
\AFFbmcc
\AFFbu
\AFFbcit
\AFFuci
\AFFcsu
\AFFcnm
\AFFduke
\AFFllr
\AFFgifu
\AFFgist
\AFFglasgow
\AFFuh
\AFFibs
\AFFicise
\AFFicl
\AFFbari
\AFFnapoli
\AFFpadova
\AFFroma
\AFFilance
\AFFkeio
\AFFkek
\AFFkcl
\AFFkobe
\AFFkyoto
\AFFliv
\AFFminn
\AFFmiyagi
\AFFnagoya
\AFFkmi
\AFFpol
\AFFnumazu
\AFFsuny
\AFFokayama
\AFFoecu
\AFFox
\AFFral
\AFFseoul
\AFFsheff
\AFFshizuokasc
\AFFsilesia
\AFFstfc
\AFFskk
\AFFtohoku
\AFFtodai
\AFFipmu
\AFFtit
\AFFtus
\AFFtoyama
\AFFtriumf
\AFFtsinghua
\AFFwu
\AFFwarwick
\AFFwinnipeg
\AFFynu

\AFFbcit
\AFFuci
\AFFcsu
\AFFcnm
\AFFduke
\AFFllr
\AFFgifu
\AFFgist
\AFFglasgow
\AFFuh
\AFFibs
\AFFicise
\AFFicl
\AFFbari
\AFFnapoli
\AFFpadova
\AFFroma
\AFFilance
\AFFkeio
\AFFkek
\AFFkcl
\AFFkobe
\AFFkyoto
\AFFliv
\AFFminn
\AFFmiyagi
\AFFnagoya
\AFFkmi
\AFFpol
\AFFsuny
\AFFokayama
\AFFoecu
\AFFox
\AFFral
\AFFseoul
\AFFsheff
\AFFshizuokasc
\AFFsilesia
\AFFstfc
\AFFskk
\AFFtohoku
\AFFtodai
\AFFipmu
\AFFtit
\AFFtus
\AFFtoyama
\AFFtriumf
\AFFtsinghua
\AFFwu
\AFFwarwick
\AFFwinnipeg
\AFFynu

\author[0000-0003-4408-6929]{F.~Nakanishi}
\email[show]{nakanishi-suv@s.okayama-u.ac.jp}  
\AFFokayama
\author{K.~Abe}
\email{sk-collab@example.com}
\AFFicrr
\AFFipmu
\author{S.~Abe}
\email{sk-collab@example.com}
\AFFicrr
\author[0000-0001-6440-933X]{Y.~Asaoka}
\email{sk-collab@example.com}
\AFFicrr
\author[0000-0003-3273-946X]{M.~Harada}
\email{sk-collab@example.com}
\AFFicrr
\author[0000-0002-8683-5038]{Y.~Hayato}
\email{sk-collab@example.com}
\AFFicrr
\AFFipmu
\author[0000-0003-1229-9452]{K.~Hiraide}
\email{sk-collab@example.com}
\AFFicrr
\AFFipmu
\author[0000-0002-8766-3629]{K.~Hosokawa}
\email{sk-collab@example.com}
\AFFicrr
\author[]{T. H. Hung}
\email{sk-collab@example.com}
\AFFicrr
\author[0000-0002-7791-5044]{K.~Ieki}
\email{sk-collab@example.com}
\AFFicrr
\AFFipmu
% ...
]
\AFFskk
\author[0000-0002-2719-2079]{J.~W.~Seo}
\email{sk-collab@example.com}
\AFFskk
\author[0000-0003-1567-5548]{I.~Yu}
\email{sk-collab@example.com}
\AFFskk

\author[0000-0003-4136-2086]{Y. Ashida}
\email{sk-collab@example.com}
\AFFtohoku
\author[0000-0002-1009-1490]{A.~K.~Ichikawa}
\email{sk-collab@example.com}
\AFFtohoku
\author[0000-0003-3302-7325]{K.~D.~Nakamura}
\email{sk-collab@example.com}
\AFFtohoku


\author{S.~Goto}
\email{sk-collab@example.com}
\AFFtodai
\author{H. Hayasaki}
\email{sk-collab@example.com}
\AFFtodai
\author{S.~Kodama}
\email{sk-collab@example.com}
\AFFtodai
\author[]{Y. Kong}
\email{sk-collab@example.com}
\AFFtodai
\author{Y. Masaki}
\email{sk-collab@example.com}
\AFFtodai
\author{Y.~Mizuno}
\email{sk-collab@example.com}
\AFFtodai
\author{T.~Muro}
\email{sk-collab@example.com}
\AFFtodai
\author[0000-0001-8393-1289]{K.~Nakagiri}
\email{sk-collab@example.com}
\AFFtodai
\author[0000-0002-2744-5216]{Y.~Nakajima}
\email{sk-collab@example.com}
\AFFtodai
\AFFipmu
\author{N.~Taniuchi}
\email{sk-collab@example.com}
\AFFtodai
\author[0000-0003-2742-0251]{M.~Yokoyama}
\email{sk-collab@example.com}
\AFFtodai
\AFFipmu

\author[0000-0002-0741-4471]{P.~de Perio}
\email{sk-collab@example.com}
\AFFipmu
\author[0000-0002-0281-2243]{S.~Fujita}
\email{sk-collab@example.com}
\AFFipmu
\author[0000-0002-0154-2456]{C.~Jes\'us-Valls}
\email{sk-collab@example.com}
\AFFipmu
\author[0000-0002-5049-3339]{K.~Martens}
\email{sk-collab@example.com}
\AFFipmu
\author[0000-0002-5172-9796]{Ll.~Marti}
\email{sk-collab@example.com}
\AFFipmu
\author{K.~M.~Tsui}
\email{sk-collab@example.com}
\AFFipmu
\author[0000-0002-0569-0480]{M.~R.~Vagins}
\email{sk-collab@example.com}
\AFFipmu
\AFFuci
\author[0000-0003-1412-092X]{J.~Xia}
\email{sk-collab@example.com}
\AFFipmu

\author[0000-0001-8858-8440]{M.~Kuze}
\email{sk-collab@example.com}
\AFFtit
\author[0000-0002-0808-8022]{S.~Izumiyama}
\email{sk-collab@example.com}
\AFFtit
\author[0000-0002-4995-9242]{R.~Matsumoto}
\email{sk-collab@example.com}
\AFFtit

\author{R.~Asaka}
\email{sk-collab@example.com}
\AFFtus
\author{M.~Ishitsuka}
\email{sk-collab@example.com}
\AFFtus
\author{M.~Sugo}
\email{sk-collab@example.com}
\AFFtus
\author{M.~Wako}
\email{sk-collab@example.com}
\AFFtus
\author[0009-0000-0112-0619]{K.~Yamauchi}
\email{sk-collab@example.com}
\AFFtus

\author[0000-0003-1572-3888]{Y.~Nakano}
\email{sk-collab@example.com}
\AFFtoyama

\author{F.~Cormier}
\email{sk-collab@example.com}
\AFFkyoto
\author{R.~Gaur}
\email{sk-collab@example.com}
\AFFtriumf
\author{M.~Hartz}
\email{sk-collab@example.com}
\AFFtriumf
\author{A.~Konaka}
\email{sk-collab@example.com}
\AFFtriumf
\author{X.~Li}
\email{sk-collab@example.com}
\AFFtriumf
\author[0000-0003-1273-985X]{B.~R.~Smithers}
\email{sk-collab@example.com}
\AFFtriumf

\author[0000-0002-2376-8413]{S.~Chen}
\email{sk-collab@example.com}
\AFFtsinghua
\author{Y.~Wu}
\email{sk-collab@example.com}
\AFFtsinghua
\author[0000-0001-5135-1319]{B.~D.~Xu}
\email{sk-collab@example.com}
\AFFtsinghua
\author{A.~Q.~Zhang}
\email{sk-collab@example.com}
\AFFtsinghua
\author{B.~Zhang}
\email{sk-collab@example.com}
\AFFtsinghua

\author{H.~Adhikary}
\email{sk-collab@example.com}
\AFFwu
\author{M.~Girgus}
\email{sk-collab@example.com}
\AFFwu
\author{P.~Govindaraj}
\email{sk-collab@example.com}
\AFFwu
\author[0000-0002-5154-5348]{M.~Posiadala-Zezula}
\email{sk-collab@example.com}
\AFFwu
\author{Y.~S.~Prabhu}
\email{sk-collab@example.com}
\AFFwu

\author{S.~B.~Boyd}
\email{sk-collab@example.com}
\AFFwarwick
\author{R.~Edwards}
\email{sk-collab@example.com}
\AFFwarwick
\author{D.~Hadley}
\email{sk-collab@example.com}
\AFFwarwick
\author{M.~Nicholson}
\email{sk-collab@example.com}
\AFFwarwick
\author{M.~O'Flaherty}
\email{sk-collab@example.com}
\AFFwarwick
\author{B.~Richards}
\email{sk-collab@example.com}
\AFFwarwick

\author{A.~Ali}
\email{sk-collab@example.com}
\AFFwinnipeg
\AFFtriumf
\author{B.~Jamieson}
\email{sk-collab@example.com}
\AFFwinnipeg

\author[0000-0001-9555-6033]{C.~Bronner}
\email{sk-collab@example.com}
\AFFynu
\author{D.~Horiguchi}
\email{sk-collab@example.com}
\AFFynu
\author[0000-0001-6510-7106]{A.~Minamino}
\email{sk-collab@example.com}
\AFFynu
\author{Y.~Sasaki}
\email{sk-collab@example.com}
\AFFynu
\author{R.~Shibayama}
\email{sk-collab@example.com}
\AFFynu
\author{R.~Shimamura}
\email{sk-collab@example.com}
\AFFynu


\collaboration{248}{The Super-Kamiokande Collaboration}


\begin{abstract}
In 2024, a failed supernova candidate, M31-2014-DS1, was reported in the Andromeda galaxy (M31), located at a distance of approximately 770 kpc.
In this paper, we search for neutrinos from this failed supernova using data from Super-Kamiokande (SK).
Based on the estimated time of black hole formation inferred from optical and infrared observations, we define a search window for neutrino events in the SK data.
Using this window, we develop a dedicated analysis method for failed supernovae and apply it to M31-2014-DS1, by conducting a cluster search using the timing and energy information of candidate events.
No significant neutrino excess is observed within the search region.
Consequently, we place an upper limit on the electron antineutrino luminosity from M31-2014-DS1 and discuss its implications for various failed SN models and their neutrino emission characteristics.
Despite the 18 MeV threshold adopted to suppress backgrounds, the search remains sufficiently sensitive to constrain the Shen-TM1 EOS, yielding a 90\% confidence level upper limit of $1.76\times10^{53}$ erg on the electron antineutrino luminosity, slightly above the expected value of $1.35\times10^{53}$ erg.
\end{abstract}



\section{Introduction}
In 2024, M31-2014-DS1, a failed supernova (failed SN) candidate---stellar deaths in which the explosion is unsuccessful and the star collapses into a black hole without a bright optical display---was reported in M31 at a distance of about $770~\rm{kpc}$~\citep{M31-2014-ds1_arXiv}.
The object exhibited a $50\%$ increase in mid-infrared (MIR) flux over a two-year period starting in 2014~\citep{2010Wright, 2014Mainzer}, while remaining undetected in optical and near-infrared (NIR) imaging observations as of 2023.
The progenitor is estimated to have been a massive hydrogen-depleted supergiant, with an estimated mass of around $20M_{\odot}$, for which models indicate that the explosion may fail~\citep{Sukhbold2016}.
In addition, the radius of the inner shell was found to decrease more than 1000 days after the initial MIR brightening.
These observations suggest that a black hole was likely to be formed sometime between 2014 and 2017.

In the final stage of stellar evolution, massive stars develop iron cores through successive fusion processes.
Stars exceeding $8M_{\odot}$ undergo core collapse, resulting in core-collapse supernovae (CCSNe).
In an archetypal CCSNe the initial implosion rebounds, ejecting the star's outer layers and leaving behind a neutron star.
However, in some cases---especially in more massive stars---the shock wave does not revive, due to the infall of the surrounding stellar material.
The star then collapses into a black hole, with minimal electromagnetic emission.

Both successful and failed SNe release more than 99\% of their gravitational binding energy in the form of neutrinos, making them a powerful probe of stellar core physics~\citep[see, e.g.,][]{1987Burrows,2006Kotake,2013Nakazato,2018Horiuchi_Kneller,2018Takiwaki_Kotake,2021Nagakura, 2021Mori}.
In the case of a successful SN, the emission of neutrinos is expected to last for several tens of seconds up to about 100 seconds. 
The luminosity of neutrinos decreases as they carry away the energy of the proto-neutron star, and it cools down.
In contrast, failed SNe are expected to exhibit short-lived neutrino emission, lasting only up to a few seconds.
This behavior arises from the formation of a black hole at the core, after which neutrinos can no longer escape.
The electron antineutrino luminosity tends to increase only until it is cut off by black hole formation.~\citep{Liebendorfer2004,Kuroda2023}.

CCSNe are commonly observed in various electromagnetic bands, such as optical and infrared bands, and extensive catalogs of such events are maintained by public databases like the Transient Name Server (TNS;~\cite{TNS2021}) and the Astronomer's Telegram (ATel;~\cite{ATeL1998}).
SNe occurring within the sensitive range of current neutrino telescopes, which typically extends only to our galaxy, are very rare.
The estimated rate is about one every several decades~\citep{Tammann1994}.

In contrast to CCSNe, the confirmed observations of failed SNe are rare, with only a few candidates reported.
For instance, a failed SN candidate was identified using the Large Binocular Telescope (LBT;~\cite{2014Hill}) data in the survey by~\cite{2015Gerke}.
The survey monitored 27 galaxies within 10 Mpc over a period of seven years, searching for massive stars that disappeared without a bright optical SN.
Among the monitored stars, one candidate whose mass is estimated to be $18-25M_{\odot}$ in NGC 6946 showed a steady decline in luminosity and vanished from the first to the last observation.
This suggests that it may have experienced a failed SN.

The reported observation on M31-2014-DS1 has drawn attention to the possibility of detecting neutrinos from M31-2014-DS1.
The prediction of the expected neutrino emission and comparison of these predictions with observational limits from Super-Kamiokande (SK) are discussed in \cite{2025Suwa}.
In that study, several nuclear equations of state (EOS), including Lattimer \& Swesty~\citep{1991Lattimer}, Shen~\citep{1998Shen,2020Shen}, Togashi~\citep{2017Togashi}, and SFHo~\citep{2013Steiner} models, are examined to evaluate how the emission depends on neutron-star properties and nuclear-physics uncertainties.
The expected neutrino detection events at the SK detector, assuming a failed SN at the distance of M31, are estimated.
However, these predictions in~\cite{2025Suwa} do not fully take into account aspects of neutrino detection such as signal efficiency and the energy threshold effect.

In this paper, we present the result of a search for neutrinos from M31-2014-DS1 using data from the SK detector.
Section~\ref{cap:super-k} describes the details of supernova neutrino observation in SK, and Section~\ref{cap:method} explains the analysis method.
The results of the present search, including the derived upper limits on the electron antineutrino luminosity, are presented in Section~\ref{cap:result}.
Finally, Section~\ref{cap:summary} summarizes the study and discusses its future prospects.

\section{Supernova neutrino observation in SK}
\label{cap:super-k}
Supernova neutrinos have been observed only once, from SN1987A in the Large Magellanic Cloud.
A total of 24 neutrino events were detected by Kamiokande~\citep{Hirata1987}, IMB~\citep{Bionta1987}, and Baksan~\citep{1988Alexeyev}, demonstrating these detectors are highly effective for observing supernova neutrinos.

SK is a large water-Cherenkov detector experiment located 1000 m underground in Kamioka, Japan~\citep{2003Fukuda}.
It contains 50,000 tonnes of ultrapure water, of which 22,500 tonnes are used as the fiducial volume.
In the present analysis, we use data from the pure-water phase, prior to the addition of gadolinium sulfate~\citep[i.e., before the SK-Gd phase][]{2004Beacom,2022bAbe}. 
This represents the most stable operation period.
The energy threshold is $3.49~\rm{MeV}$ of electron kinetic energy~\citep{Solar_sk4_2016}.

SK is highly sensitive to neutrinos from CCSNe in the Milky Way.
For a CCSN at a distance of 10 kpc, about $\mathcal{O}(1000)$ inverse beta decay (IBD; $\bar{\nu}_e + p \rightarrow e^+ + n$) events are expected to be observed.
In IBD interactions, the prompt positron emits Cherenkov light, while the delayed neutron is captured on a proton, producing a $2.2~\rm{MeV}$ gamma ray.
However, this delayed signal is often undetectable in the pure-water phase due to the low detection efficiency.
Taking into account the detection efficiency, the expected number of detected neutrino events from CCSNe occurring in nearby galaxies (e.g., within 1~Mpc) is significantly reduced to $\mathcal{O}(0.1)$. 
Since supernova neutrinos are emitted over a timescale of several to tens of seconds, the few events expected from such distances would appear as a small cluster of events in the SK data.
Therefore, we adopt a time-clustering analysis to search for such event grouping.

SK has conducted several searches for SN burst neutrinos~\citep{2007Ikeda, 2022Mori}.
These analyses scanned the full dataset with multiple cluster definitions reflecting theoretically motivated timescales, including the initial collapse and bounce, shock revival, and proto-neutron star cooling.
The searches are not directed at known astrophysical objects, but rather aimed to identify SNe in a blind manner, including those that might have occurred but remained undetected in optical observations due to dust obscuration.
A search of the 2008-2018 dataset likewise found no burst-like clusters~\citep{2022Mori}.
In the present analysis, we specifically applied a cluster search criterion optimized for M31-2014-DS1, as described in Section~\ref{cap:cluster}, representing a more targeted approach than previous blind searches.

\section{Analysis method}
\label{cap:method}
We performed a time-clustering analysis of SK data to search for neutrino signals potentially associated with the failed SN candidate M31-2014-DS1.
Since supernova neutrinos are emitted over a timescale of a few seconds, a signal from M31 would likely appear as a small cluster of temporally correlated events.
We defined a 10-second time window and cluster criteria, and then evaluated the expected background to determine the optimal energy threshold.
The 10-second window is motivated by numerical simulations of failed SNe, which predict black hole formation within a few seconds after core bounce. 
At the distance of M31, the expected number of detected neutrinos in SK is $\mathcal{O}(1)$. 
The estimated background rate corresponds to approximately $1\times10^{-5}$ events per 10-second window above the $18~\rm{MeV}$ threshold, indicating that the observation of two or more events within such a time window would represent a statistically significant excess over background expectations.
The event selection applied in this search is described in Section~\ref{cap:cluster}.

\subsection{Determination of the search region}
\begin{figure*}[htbp]
    \centering
    \includegraphics[scale=0.9]{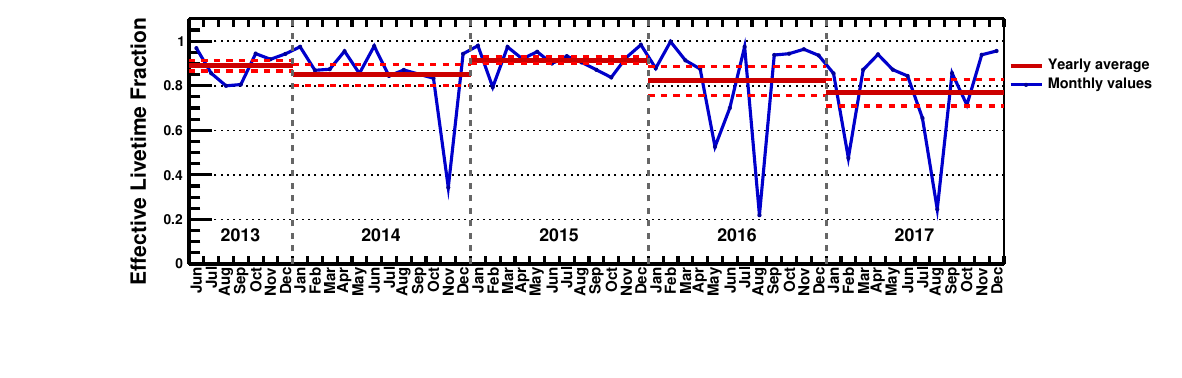}
    \caption{Livetime fraction of SK used in this analysis. The blue line indicate the effective livetime fraction for each month, while the red solid lines show the yearly averaged effective livetime fraction. The red dashed lines represent the standard deviation of the yearly mean. The loss of livetime is mainly attributed to detector calibrations and maintenance periods.}
    \label{fig:livetime}
\end{figure*}

The exact determination of black hole formation time is difficult with optical telescope observations, because the stellar core where the black hole forms is not directly visible due to the surrounding stellar envelope.
Therefore, we defined the ``signal time range'' as the period fully encompassing the expected timeframe for black hole formation, and the ``background time range'' as the periods immediately before and after it to estimate the background rate. The details are summarized in Table~\ref{tab:time_ranges}.
The effective livetime fraction during the signal time range was evaluated. 
Although the possibility that the SK data were unavailable at the time of the M31-2014-DS1 event (e.g., due to calibration or maintenance periods) cannot be completely excluded, the yearly averaged livetime efficiencies were evaluated as shown in Figure~\ref{fig:livetime}. 
The efficiencies were 89.1~$\pm$~2.4\%, 84.9~$\pm$~4.6\%, 91.5~$\pm$~1.6\%, 82.2~$\pm$~6.4\%, and 76.8~$\pm$~6.0\% for 2013, 2014, 2015, 2016, and 2017, respectively, indicating that the detector was generally operational during this period.
For the background time range, the efficiency were evaluated to be 88.2~$\pm$~3.6\% before and 77.4~$\pm$~3.1\% after the signal time range.

\begin{deluxetable}{lcc}
\tablecaption{Definition of the signal and background time ranges.\label{tab:time_ranges}}
\tablehead{
\colhead{Time range} & \colhead{Period} & \colhead{Duration (days)}}
\startdata
Signal time range   & 2013 Jun. 1 -- 2017 Dec. 31 & 1416.3 \\
Background time range (before) & 2012 Mar. 15 -- 2013 Jun. 1 & 390.5\\
Background time range (after)  & 2018 Jan. 1 -- 2018 May 31 & 118.1\\
\enddata
\end{deluxetable}
The pre-signal background time range is defined as the interval following a known supernova candidate, SN2012aw, observed in March 2012~\citep{2012ATel}, and extending up to the beginning of the signal time range.
The post-signal background time range is defined until the end of the pure-water phase after the signal time range.

\subsection{Event selection}
\label{cap:selection}
The IBD reaction is the target of the diffuse supernova neutrino background (DSNB) search in the neutrino detector; therefore, the event selection and reconstruction follow the previous DSNB search in the pure-water phase~\citep{2021PhRvD_DSNBsk4}.
The event selection requires a reconstructed electron or positron total energy above $8~\rm{MeV}$ to avoid entering radioactive background events into the data sample.
However, to secure the signal event, we did not apply neutron tagging in this search.

In SK, the main background events in the SN neutrino energy region are radioimpurities, spallation events, decay electrons from muons, and atmospheric neutrinos.
In the case of radioimpurities, backgrounds mainly originate from the radio impurities dissolved in purified water~\citep{Nakano2020}, the material of the detector structure, and the surrounding rock of the SK tank.
Such backgrounds can be reduced by a fiducial volume (FV) cut, which excludes events occurring within $2~\rm{m}$ of the tank wall.
As a result, the tank volume is reduced from the entire inner detector (ID) volume of $32,500~\rm{tonnes}$ to $22,500~\rm{tonnes}$.

A ``spallation event'' originates from the decays of unstable isotopes produced when atmospheric muons interact with oxygen nuclei in water~\citep{Li2014,2015Li,2015Li2}.
The primary spallation process generates hadrons such as pions and neutrons, while the subsequent nuclear interactions produce a variety of beta decay nuclei.
The final observables, including beta and gamma rays, have energies from a few MeV up to $\sim20~\rm{MeV}$.
Such spallation events are removed by a spallation cut, which considers pairs of two low-energy events passing the preselection, including information on muons that precede low-energy events within 30 seconds~\citep{Locke2024,Zhang2016}.
An additional cut removes events within 4~m and 60~s of another low-energy event, to suppress correlated spallation backgrounds.
Details can be found in~\cite{2021PhRvD_DSNBsk4}.

The dominant background above $20~\rm{MeV}$ originates from charged-current quasi-elastic (CCQE) interactions of atmospheric neutrinos.
However, constraints on the CCQE event rate are limited by statistics in the background time range, making it difficult to estimate the background events.
Therefore, as described in the next section, we estimated the CCQE background using a simulation.

\subsection{Background event estimation}
\label{cap:bkg_estimation}
To reliably estimate the background rate, we use an atmospheric neutrino Monte Carlo (MC) simulation corresponding to 500 years based on the HKKM 2011 atmospheric neutrino flux~\citep{2007Honda,2011Honda} and the NEUT 5.3.6 neutrino interaction simulator~\citep {2009Hayato}.
The flux of atmospheric neutrino CCQE events in the simulation is normalized by fitting the reconstructed energy spectrum in the range $30 < E_{\rm{rec}} < 60~\rm{MeV}$ to the data.
The fit employed the full $\sim$10-year data set, including the signal time range, in order to ensure sufficient statistics.
Although the signal time range is included, this energy region is above the typical expected neutrino energies from failed SN, and even if a few signal events were present, their contribution to the spectrum would be negligible.

As a result, the estimated background rate is $10.58~\mathrm{day^{-1}}$ for $8~\mathrm{MeV} < E_{\rm{rec}} \leq 80~\mathrm{MeV}$ and $0.09~\mathrm{day^{-1}}$ for $18~\mathrm{MeV} < E_{\rm{rec}} \leq 80~\mathrm{MeV}$.

\subsection{Evaluation of background cluster}
\label{cap:cluster}

\begin{figure}[htbp]
    \centering
    \includegraphics[scale=0.42]{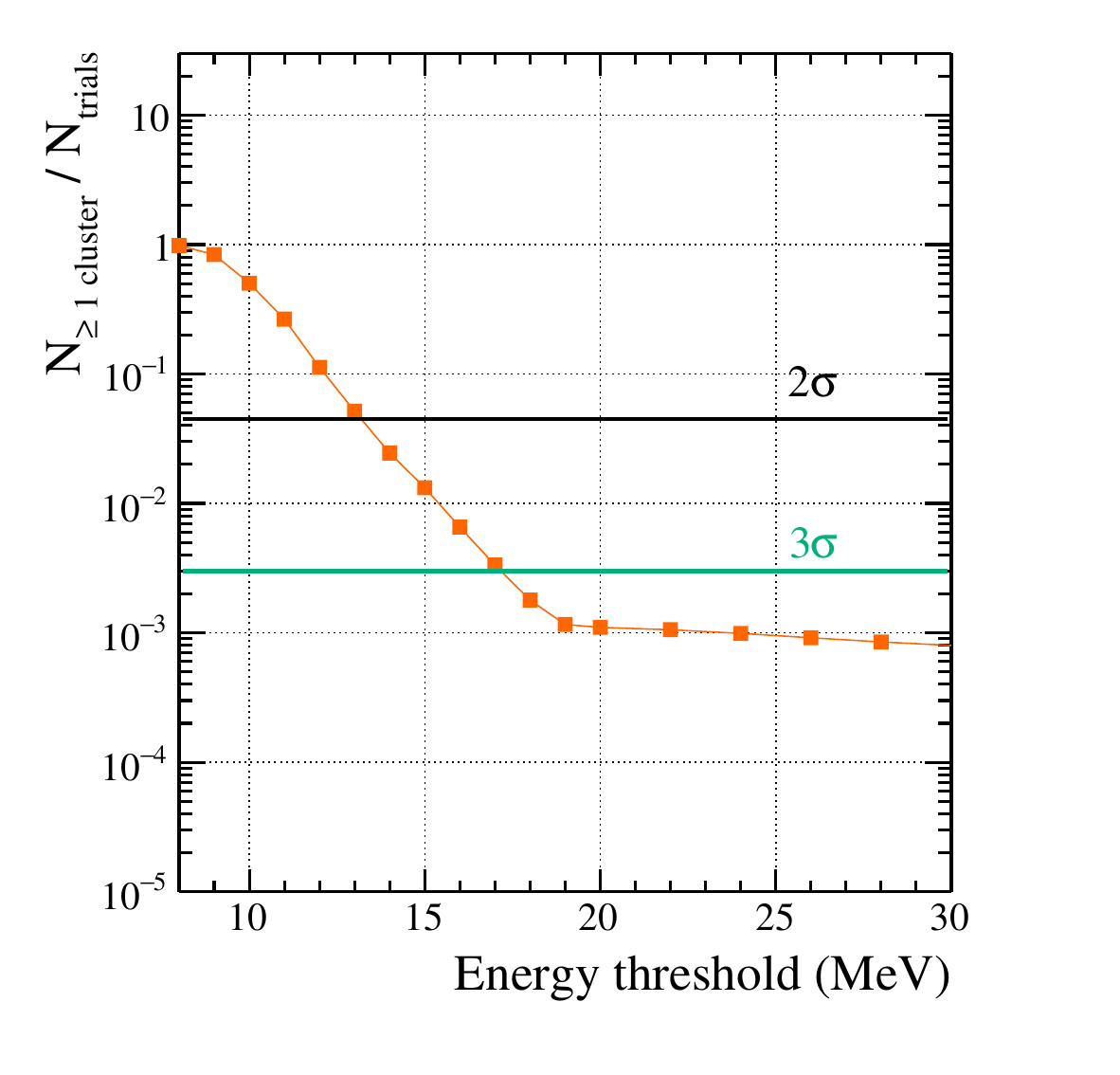}
    \caption{The expected probability of observing clusters due to background events at each energy threshold. The red points indicate the probability of observing one or more clusters within the signal time range. The black and green lines represent the fractions of trials without any clusters at the $2\sigma$ and $3\sigma$ levels, respectively.} 
    \label{fig:cls_prob}
\end{figure}

To determine the optimal positron energy threshold for the cluster search, we estimate the background event rate and probability of background-induced clusters.
The total background estimate consists of two components: (1) a data-driven component estimated by the background in SK below $20~\rm{MeV}$ during the background time range, as summarized in Table~\ref{tab:time_ranges}, and (2) an MC-based atmospheric neutrino background above $20~\rm{MeV}$, as described in Section~\ref{cap:bkg_estimation}.
We calculated the cluster probability using a toy MC approach.
For each energy threshold, we simulated $10^7$ trials, where background events are randomly distributed according to the estimated rate.
The background events are expected to follow a nearly random distribution.
Although some correlation between atmospheric neutrinos and spallation products may exist due to their association with atmospheric muons, such effects are not expected to significantly impact the results of this study.
In each trial, we searched for clusters defined as $\geq2$ events within a 10-second window, and counted the number of such occurrences.
The cluster probability is defined as the fraction of trials that contain at least one cluster:
\begin{equation}
    P = \frac{N_{\ge 1\,{\rm cluster}}}{N_{\rm trials}},
\end{equation}
where $N_{\ge 1\,{\rm cluster}}$ is the number of trials with one or more clusters, and $N_{\rm trials}$ is the total number of trials.
We require the probability to be below the $3\sigma$ level ($P < 0.003$), ensuring that any observed cluster is not likely to be due to statistical fluctuations of the background.
As shown in Figure~\ref{fig:cls_prob}, the cluster probability decreases with increasing energy threshold.
Based on this evaluation, we set the energy threshold to 18 MeV, which satisfies the $3\sigma$ criterion.
At this threshold, the observation of two or more events within 10~seconds would constitute a detection.

\section{Result}
\label{cap:result}
Following the cluster search procedure, no clusters consisting of two or more events within a 10-second window were found in the signal time range.

Based on this result, we set the upper limit on the electron antineutrino luminosity using two complementary methods.
In both methods, the expected number of signal events in SK is estimated by taking into account the detector response and signal efficiency.
In the first approach, we derive an upper limit on the neutrino luminosity by calculating the $90\%$ confidence level upper limit on the neutrino fluence at SK, based on the assumption of specific failed SN models.
The fluence upper limit is calculated as~\citep{GW2016}:
\begin{equation}
\label{equ:fluence}
\Phi = 
\frac{N_{90}}{
    N_T \int dE_\nu\, \lambda(E_\nu) \sigma(E_\nu) R(E_e, E_{\text{vis}}) \epsilon(E_{\text{vis}})
}
\end{equation}
Here, $\Phi$ is the upper limit on the neutrino flux.
$N_{90}$, $E_e$, $E_{\text{vis}}$ represent the $90\%$ confidence level (C.L.) upper limit on the number of neutrino events in the search window, the true electron energy, the reconstructed visible energy, respectively.
Since no clusters are observed within a 10-second time window, we set $N_{90}=3.89$, based on the probability to observe at least 2 events.
$N_T$ is the number of target particles in the detector.
$\lambda(E_\nu)$ denotes the normalized neutrino energy spectrum,
$\sigma(E_\nu)$ is the neutrino interaction cross section from Strumia-Vissani model~\citep{2003Strumia},
$R(E_e, E_{\text{vis}})$ is the energy response function that connects $E_e$ to $E_{\text{vis}}$,
and $\epsilon(E_{\text{vis}})$ is the detection efficiency as a function of visible energy.
The effect of $R(E_e, E_{\text{vis}})$ is expected to be small in the energy range relevant to this analysis, and is therefore not taken into account in this analysis.
Using the fluence upper limit $\Phi$, we derive a corresponding upper limit on the total emitted electron antineutrino luminosity under the assumption of isotropic emission.
The relation is given by:
\begin{equation}
L_{\nu} = 4\pi D^2 \langle E_{{\bar{\nu}_e}} \rangle \cdot \Phi,
\end{equation}
where $D = 770~\rm{kpc}$ and $\langle E_{\bar{\nu}_e} \rangle = \int^\infty_0 \lambda(E)dE$ is the mean energy of the emitted electron antineutrinos based on the failed SN model spectrum~\citep {2021Nakazato,2007Sumiyoshi,2008Sumiyoshi,2025Choi}.
Although neutrinos of all flavors are emitted in a CCSN, only the detectable electron antineutrinos are taken into account in this calculation.
\begin{figure}[htbp]
    \centering
    \includegraphics[scale=0.165]{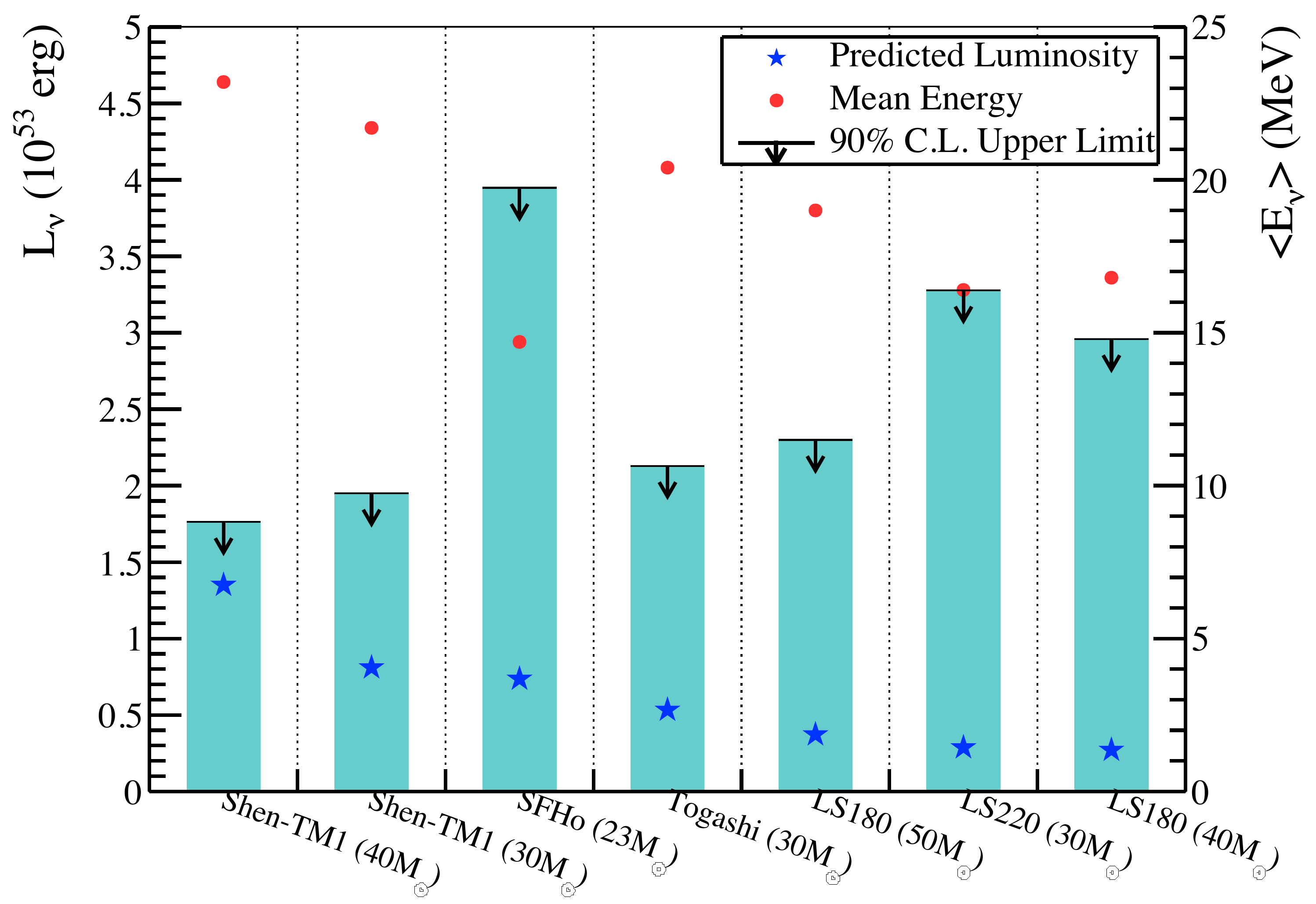}
    \caption{Upper limits on the neutrino luminosity for each failed SN model. The models are arranged from left to right in order of decreasing predicted luminosity. The star marks the predicted luminosity of the model, while the black downward arrow indicates the $90\%$ C.L. upper limit obtained in the present work. The circle points show the average energy for each model, plotted with the right vertical axis, while the luminosity is shown with the left axis. From left to right, the models correspond to~\cite {2021Nakazato,2007Sumiyoshi,2008Sumiyoshi,2025Choi}.}
    \label{fig:lumi_upper}
\end{figure}
In Figure~\ref{fig:lumi_upper}, the luminosity upper limits for each failed SN model are shown.
For models such as Shen-TM1($40M_{\odot}$), which predict both high luminosity and higher mean neutrino energy, the upper limit derived from our analysis approaches the predicted neutrino luminosity. 
In contrast, for models like SFHo ($23M_{\odot}$), which have lower mean neutrino energy, the fluence upper limit is more affected by the analysis energy threshold, resulting in upper limits that tend to be farther from the model predictions.

\begin{figure*}[htbp]
    \centering
    \includegraphics[scale=0.65]{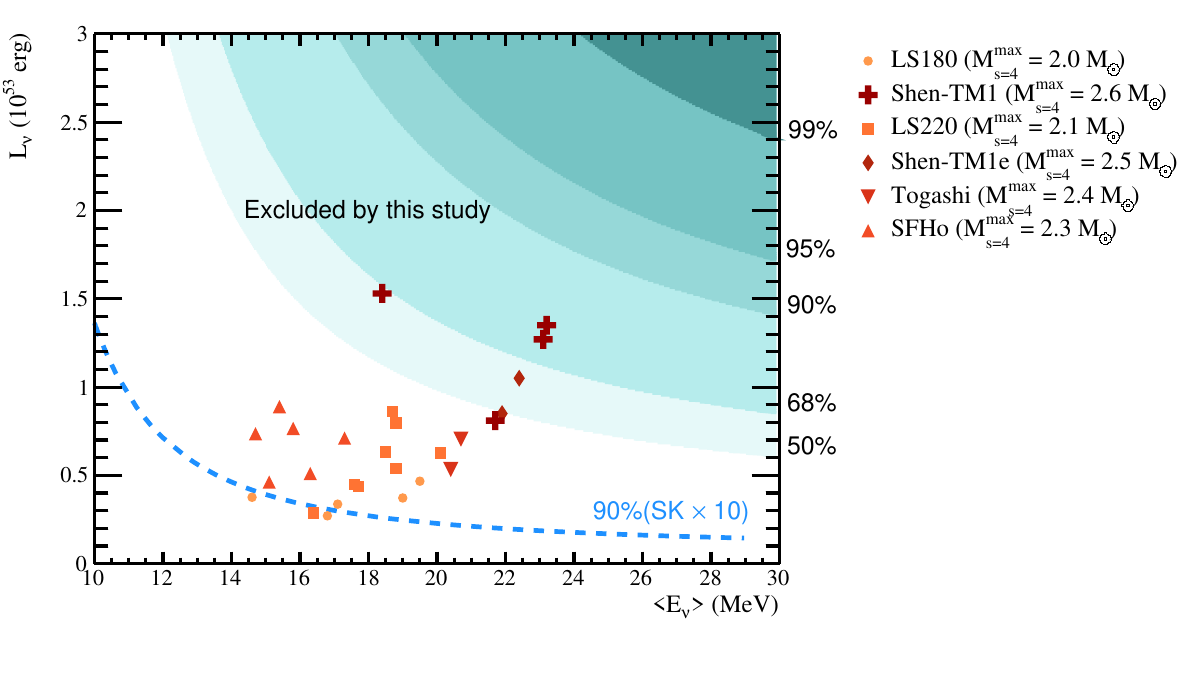}
    \caption{Electron antineutrino luminosity $L_{\nu}$ versus mean energy $\langle E_{\bar{\nu}_e} \rangle$. Shaded regions (light to dark cyan) represent the probability contours (50\%, 68\%, 90\%, 95\%, and 99\%) for detecting at least two correlated events in SK, based on Poisson statistics for a source located at 770 kpc. The regions above these bands are excluded based on the nondetection of time-clustered events in SK. Filled markers show individual simulation results for six nuclear equations of state (EOS): LS180, LS220, SFHo, Togashi, Shen-TM1e, and Shen-TM1. Each model corresponds to Figure 1 and Table 1 in \cite{2025Suwa}. The blue dashed line indicates the 90\% probability contour for a detector with ten times the fiducial mass of SK, assuming Poisson statistics for a source at the distance of $770~\rm{kpc}$.}
    \label{fig:lumi_mean}
\end{figure*}

In another approach, we estimated the number of expected events for IBD using a Fermi--Dirac neutrino spectrum characterized by a temperature $T$ and a total electron antineutrino luminosity $L_{\nu}$.
The expected number of events is calculated as:
\begin{equation}
    N_{\rm{event}} = \int^{\infty}_{E_{\rm{th}}} N_T \cdot \phi(E_{\nu}) \cdot \sigma(E_{
    \nu}) \cdot \epsilon(E_{\nu}) \cdot dE_{\nu}.
\end{equation}
Here, $\phi(E_{\nu})$ is the neutrino flux at the Earth, and other variables are the same as Equation~\ref{equ:fluence}.
The flux $\phi(E_{\nu})$ is derived from a Fermi--Dirac distribution assuming a $T$ and $L_{\nu}$, with zero chemical potential.
The distribution is normalized to satisfy the condition:
\begin{equation}
    L_{\nu} = \int^{\infty}_0 E_{\nu} \cdot \frac{dN}{dE_{\nu}} dE_{\nu}.
\end{equation}
The distance to the source, $D$, is assumed to be 770~kpc, and the flux is given by $\phi(E_\nu) = 1 / 4\pi D^2 \cdot dN / dE_\nu$.
Figure~\ref{fig:lumi_mean} shows the relationship between $L_{\nu}$ and the mean neutrino energy, $\langle E_{\nu}\rangle$.
The shaded regions represent the excluded parameter space derived from the nondetection of time-clustered events in SK, assuming Poisson statistics.
The contours represent confidence levels of $50\%$, $68\%$, $90\%$, $95\%$, and $99\%$, respectively, which translates to a Poisson upper limit of 1.68, 2.35, 3.89, 4.74, and 6.64 events in the search window.
Contours indicate combinations of emission parameters ($L_{\nu}$, $\langle E_\nu\rangle$) for which the probability of detecting two or more events would have exceeded the given confidence level.
While the $18~\rm{MeV}$ threshold in the cluster search was chosen to reduce the number of backgrounds effectively, the sensitivity remains sufficient to constrain the Shen-TM1 relativistic mean-field EOS, which is relatively stiff and yields a maximum cold neutron star mass of $2.2M_{\odot}$, at a confidence level exceeding $68\%$, demonstrating that this method can effectively constrain optimistic black hole formation scenarios.

As a notional future scenario, we consider a detector with a fiducial volume ten times larger than that of SK, representative of the scale of Hyper-Kamiokande~(HK, \cite{2018HyperK}).
In this estimation, we adopt the same analysis conditions as in this study, such as the signal efficiency and the positron energy threshold.
In Figure~\ref{fig:lumi_mean}, the blue dashed line indicates the upper limit at 90\% C.L. based on the Poisson expectation when no clusters are observed.
Under this assumption, many failed SN models can be constrained, highlighting the significant discovery potential of future large-scale detectors. 
We note that more detailed and dedicated studies are being carried out for HK, which will provide a more realistic assessment of its sensitivity.

\section{Summary and future prospect}
\label{cap:summary}
We searched for neutrinos from M31-2014-DS1 using SK data.
The timing of black hole formation is uncertain, as the inner core is obscured in optical and infrared wavelengths.
Accordingly, the search period was defined from June 1, 2013, to December 31, 2017.
In this analysis, we developed a cluster search method optimized for M31-2014-DS1, where a failed SN at the distance of M31 is expected to produce a short neutrino burst. 
A cluster is defined as two or more events within 10 seconds, and an $18~\rm{MeV}$ positron energy threshold was applied to sufficiently reduce backgrounds.
This procedure can be applied to other failed SN searches by re-optimizing the energy threshold for the search based on the telescope's observation period and the cluster criteria inferred from the expected number of events at a given distance.
In the case of M31-2014-DS1, no cluster satisfying the selection criteria was found in the signal time range.

From the result, we derive upper limits on the neutrino luminosity using two spectral assumptions: failed SN model predictions and a Fermi--Dirac distribution.
For failed SN models, the limits depend on the spectral characteristics.
Models such as Shen-TM1 ($40M_{\odot}$) with high luminosity and mean energy yield limits close to the predicted energy, whereas models like SFHo ($23M_{\odot}$) with lower mean energy are more affected by the $18~\rm{MeV}$ threshold, showing larger deviations.
In the Fermi--Dirac-based analysis, models with lower average neutrino energy require higher luminosity to maintain the same expected number of detected events.
As a result, the exclusion contours shift toward higher luminosities as the assumed average energy decreases.
This behavior is reflected in the limit contours shown in Figure~\ref{fig:lumi_mean}.
Although these contours do not directly provide the numerical limits, the upper limits are evaluated separately under each EOS model assumption as shown in Figure~\ref{fig:lumi_upper}.
Following this procedure, the search demonstrates sufficient sensitivity to constrain the Shen-TM1 EOS, resulting in a 90\% C.L. upper limit of $1.76\times10^{53}$~erg on the electron antineutrino luminosity, which is slightly above the expected value of $1.35\times10^{53}$~erg.

Improved estimation of the black hole formation time would allow for a narrower signal time range and a lower energy threshold, thereby enhancing the sensitivity to a wider range of failed SN models.
In addition, future advances in optical observations may also contribute to narrowing the signal time range.
Ongoing optical surveys with facilities such as the LBT and the Subaru Telescope equipped with Hyper Suprime-Cam (HSC,~\cite{2018Miyazaki}) already contribute to searches for failed SNe.
In the future, the Large Synoptic Survey Telescope (LSST,~\cite{2019Zeljko}) will extend these efforts with a ten-year, all-sky time-domain survey.
Furthermore, upcoming neutrino detectors such as HK, JUNO~\citep{2016JUNO}, and DUNE~\citep{2020DUNE} will offer significantly improved sensitivity to extragalactic failed SNe.
These next-generation detectors are expected to detect such events and thereby place much stronger constraints on theoretical models.

\section{Acknowledgement}
We gratefully acknowledge the cooperation of the Kamioka Mining and Smelting Company. The Super-Kamiokande experiment has been built and operated from funding by the Japanese Ministry of Education, Culture, Sports, Science and Technology; the U.S. Department of Energy; and the U.S. National Science Foundation. Some of us have been supported by funds from the National Research Foundation of Korea (NRF-2009-0083526, NRF-2022R1A5A1030700, NRF-2022R1A3B1078756, RS-2025-00514948) funded by the Ministry of Science, Information and Communication Technology (ICT); the Institute for Basic Science (IBS-R016-Y2); and the Ministry of Education (2018R1D1A1B07049158, 2021R1I1A1A01042256, RS-2024-00442775); the Japan Society for the Promotion of Science; the National Natural Science Foundation of China under Grants No. 12375100; the Spanish Ministry of Science, Universities and Innovation (grant PID2021-124050NB-C31); the Natural Sciences and Engineering Research Council (NSERC) of Canada; the Scinet and Digital Research of Alliance Canada; the National Science Centre (UMO-2018/30/E/ST2/00441 and UMO-2022/46/E/ST2/00336) and the Ministry of  Science and Higher Education (2023/WK/04), Poland; the Science and Technology Facilities Council (STFC) and Grid for Particle Physics (GridPP), UK; the European Union’s Horizon 2020 Research and Innovation Programme H2020-MSCA-RISE-2018 JENNIFER2 grant agreement no.822070, H2020-MSCA-RISE-2019 SK2HK grant agreement no. 872549; and European Union's Next Generation EU/PRTR  grant CA3/RSUE2021-00559; the National Institute for Nuclear Physics (INFN), Italy.o

\clearpage
\bibliography{journal}{}
\bibliographystyle{aasjournalv7}



\end{document}